\documentclass[twocolumn,showpacs,preprintnumbers,amssymb]{revtex4}
\usepackage{pdfpages}
\usepackage{graphics}
\usepackage{epsfig} 
\usepackage{hyperref}
\usepackage{graphics}
\usepackage{color}      
\usepackage{graphicx}
\usepackage{graphicx}
\usepackage{dcolumn}
\usepackage{amsmath}

\usepackage{graphicx}
\usepackage{subcaption}

\begin{document}

\title{ Curzon–Ahlborn-Type Efficiency in a Brownian Heat Engine with Exponential Temperature Profile}
\author{Mesfin Asfaw  Taye}
\affiliation {West Los Angeles College, Science Division \\9000  Overland Ave, Culver City, CA 90230, USA}

\email{tayem@wlac.edu}

\begin{abstract}

 We investigate a Brownian heat engine wherein a particle moves through a periodic ratchet potential under an exponentially decreasing temperature profile, a spatial configuration that closely resembles experimentally realizable conditions such as laser-induced thermal gradients and thermoplasmonic heating. This model yields exact analytical expressions for the particle current, thermodynamic efficiency, entropy production, and coefficient of performance (COP), and uniquely recovers the Curzon–Ahlborn efficiency and the corresponding endoreversible COP exactly in the quasistatic limit. These findings provide a rare and rigorous realization of endoreversible thermodynamics at the mesoscopic scale because they are derived directly from microscopic stochastic dynamics without recourse to phenomenological assumptions, asymptotic approximations or coarse-graining techniques. Although the derived efficiency and COP are exact, they remain strictly below the Carnot limit, reflecting the inherent irreversibility embedded within the endoreversible framework. Furthermore, we show  that in comparison to linear and piecewise-constant temperature profile cases, the exponential temperature profile leads to significantly higher particle velocities,  higher  entropy production, but  lower thermodynamic efficiency, which   underscores the fundamental trade-off between transport speed and energy cost. We further extend our analysis to networks of interacting Brownian motors operating in spatially non-uniform thermal environments. Numerical simulations confirm our analytical predictions and reveal the critical roles of temporal dynamics and external load in shaping motor performance, as well as  transport directionality. Importantly, the exponential temperature profile is not only analytically tractable but also experimentally viable, providing a powerful platform for probing the emergence of macroscopic thermodynamic behavior from the underlying microscopic nonequilibrium dynamics.
\end{abstract}
\pacs{Valid PACS appear here}


\maketitle

\section{Introduction}

Can directed motion emerge in the absence of net external force? How do spatial variations in temperature affect the performance of microscopic engines? These questions lie at the heart of modern studies on noise-induced transport and Brownian thermodynamics, where thermal fluctuations, rather than deterministic driving, power motion at micro- and nanoscale dimensions \cite{c8,cc8,mg1,mg2,mg3,mg4,mg5,mg6,mg7,mg8}. These mechanisms have broad relevance in biological transport, microfluidic systems, and nanoscale energy conversion.

The emergence of directed motion under such non-equilibrium conditions has been extensively studied in the context of Brownian motors, which convert stochastic fluctuations into useful work. The foundational work by Reimann et al. \cite{am7} laid the groundwork for a theoretical understanding of directed transport in flashing and rocking ratchet systems. This influential contribution has stimulated a substantial body of research on Brownian heat engines that  operate  in spatially heterogeneous thermal environments \cite{c9,c10,c11,c12,c13,c14,c15,c16,c17,c18,c19,c20,c21,c22,c23}. Within this framework, considerable attention has been  given to a Brownian motor  that works  due to piecewise constant \cite{c19}, linearly decreasing \cite{c23}, and quadratically decreasing \cite{c22} temperature profile  cases. Each of these cases exhibits distinct thermodynamic characteristics and offers valuable insights into the role of thermal gradients in driving nonequilibrium transport.

In our previous studies \cite{c10,c19,c23,c22}, we analytically solved Brownian motor models subjected to these temperature landscapes. The \emph{piecewise constant} case \cite{c19}, representing alternating hot and cold thermal contacts, achieves a relatively high efficiency but with limited current. The \emph{linearly decreasing} temperature profile \cite{c23} enhances the entropy production and transport speed, but only approximates the endoreversible efficiency in the quasistatic regime. In contrast, the \emph{quadratic} profile \cite{c22} yields even higher particle velocities, albeit with reduced efficiency and increased thermodynamic irreversibility. These findings underscore the strong influence of the spatial structure of the temperature field on the performance of Brownian engines.

From an experimental standpoint, the exponentially decaying temperature profile offers significant advantages over the commonly used piecewise constant, linear, or quadratic configurations. Piecewise constant profiles require sharply defined thermal interfaces, which are difficult to realize at the micro- and nanoscale, and introduce non-physical discontinuities. In contrast, exponential gradients arise naturally from localized heating sources, such as focused laser beams or plasmonic absorbers, where the interplay of Beer–Lambert absorption and steady-state heat conduction yields smooth and spatially continuous thermal fields \cite{c25,c26,c27,c28}. These profiles are not only analytically tractable but also experimentally verified  \cite{c25,c26}. Although linear and quadratic gradients can be engineered using patterned heaters or thermally graded substrates, such approaches require complicated  fabrication and strict spatial control. Exponential gradients, by contrast, are readily implemented in fluidic systems via optical absorption, enabling the robust realization of spatially varying temperature fields for thermophoretic trapping, optothermal flow control, and nonequilibrium transport phenomena \cite{c27,c28}. Accordingly, the exponential profile provides a compelling synthesis of physical relevance, experimental accessibility, and theoretical simplicity, establishing it as a practical and realistic basis for modeling thermal driving in Brownian systems at the nanoscale.

Motivated by these insights, the present study investigates a Brownian heat engine operating under an \emph{exponentially decreasing temperature profile.} This configuration is not only analytically tractable but also closely resembles experimentally realizable systems such as laser-induced localized heating in fluids. A central and novel result of this study is that, in the quasistatic limit, the engine attains the exact efficiency of an endoreversible heat engine:
$
\eta = 1 - \sqrt{T_c / T_h}$,
as well as the corresponding coefficient of performance for refrigeration.
\begin{equation}
\mathrm{COP} = \frac{1}{\sqrt{T_h / T_c} - 1}.
\end{equation}
To the best of our knowledge, this represents the first exact realization of the Curzon–Ahlborn efficiency in a stochastic heat engine with a smooth and continuous temperature field. Previous models with linear gradients only approach this limit asymptotically \cite{c24}. This result establishes a direct link between the geometry of the thermal gradient and the fundamental limits of the thermodynamic performance.

We further support these analytical findings through Brownian dynamic simulations.  Via the simulations,  we  validate the behavior of the velocity, efficiency, and entropy-related quantities across a range of parameters. We show that Brownian particles act   as a heat engine or refrigerator depending on the external load and thermal asymmetry. We also extend our analysis to \emph{networks} of Brownian motors subjected to exponential temperature fields. Notably, while rates such as velocity and entropy production remain independent of network size, extensive quantities such as entropy scale linearly with system size. This reveals a clear separation between local dynamical features and collective thermodynamic behavior.
Finally, we compare the performance of the exponential temperature profile with that of the piecewise constant case. Although the exponential gradient yields a higher transport velocity, it is accompanied by lower efficiency and greater entropy production,  which  reflects   a trade-off between power output and thermodynamic irreversibility.

At this point, we emphasize that the exact realization of classical thermodynamic bounds, such as the Curzon–Ahlborn efficiency and the corresponding endoreversible coefficient of performance (COP) within mesoscopic models, remains remarkably rare. In this study, we demonstrate that a Brownian heat engine operating under an exponentially decreasing temperature profile attains both bounds exactly in the quasistatic regime without recourse to approximations or asymptotic limits. This result is not merely mathematically elegant; it is physically profound, as it provides a rare example of a mesoscopic system that recovers the defining performance limits of endoreversible thermodynamics. Although the Curzon–Ahlborn efficiency and its associated COP are cornerstones of finite-time thermodynamics, their original derivation relies on phenomenological arguments and effective continuum assumptions. In contrast, our model yields results from first principles through exact microscopic dynamics, offering a rigorous alternative to the traditional macroscopic formulations. Although we have previously derived exact solutions for systems with linear, quadratic, and piecewise constant temperature profiles, none of these exactly reproduce the endoreversible bounds. It is only in the case of an exponential temperature gradient that the full structure of endoreversible thermodynamics emerges naturally and completely from the microscopic level. This positions the exponential profile as an analytically privileged and physically realistic scenario, providing not only exact solutions for efficiency, current, COP, and entropy production, but also a rare and instructive bridge between stochastic thermodynamics and classical thermodynamic limits. In doing so, it offers a powerful framework that helps  examine the microscopic foundations of endoreversibility and advances our understanding of optimal energy conversion in nonequilibrium systems.

The remainder of this paper is organized as follows: Section II presents  the model. In  Section III, we  explore the dependence of the velocity, efficiency, and  coefficient  of performance of the refrigerator  on the model parameters. In Section IV,   we  examine  the dependence of thermodynamic relations, such as entropy and entropy production rate,   on the model parameters.  In Section V,  we consider a network of Brownian  motors  and study their  dynamics.  In Section VI, we study the model system using numerical simulations.  Section VII presents the summary and conclusions.

\section{Model Description}

We now consider  a Brownian  particle that  walks  along   a periodic sawtooth potential $U_s(x)$ with an external linear load $f$. The ratchet  potential  is given by 
\begin{equation} 
  U_{s}(x)=\left\{\begin{array}{cl}
   2U_{0}[{x\over L_0}],&if~ 0< x \le L_0/2;\\
   2U_{0}[{-x\over L_0}+1],&if~ L_0/2 < x \le L_0.\end{array}\right.
   \end{equation}
The ratchet potential is coupled with  a temperature that decreases linearly  along the reaction coordinate as 
\begin{equation}
T(x) = T_h e^{-\alpha x / L_0}, \quad 0 \leq x < L_0,
\end{equation}
where
\begin{equation}
 \alpha = \ln(T_h/T_c).
\end {equation} The  temperature  decays   exponentially from the hot temperature $T_h$  at $x=0$ to cold temperature   $T_c$  at $x = L_0$  (see Fig. 1).

For  piecewise constant temperature case
 \begin{equation}
T(x)=\left\{
\begin{array}{ll}
T_{h},& \text{if} ~~~0 \le x \le {L_{0}\over 2};\\
T_{c},& \text{if} ~~~ {L_{0}\over 2} \le x \le L_{0}.
\end{array}
\right.
\end{equation}

\begin{figure}[ht]
\centering
{
    \includegraphics[width=12cm]{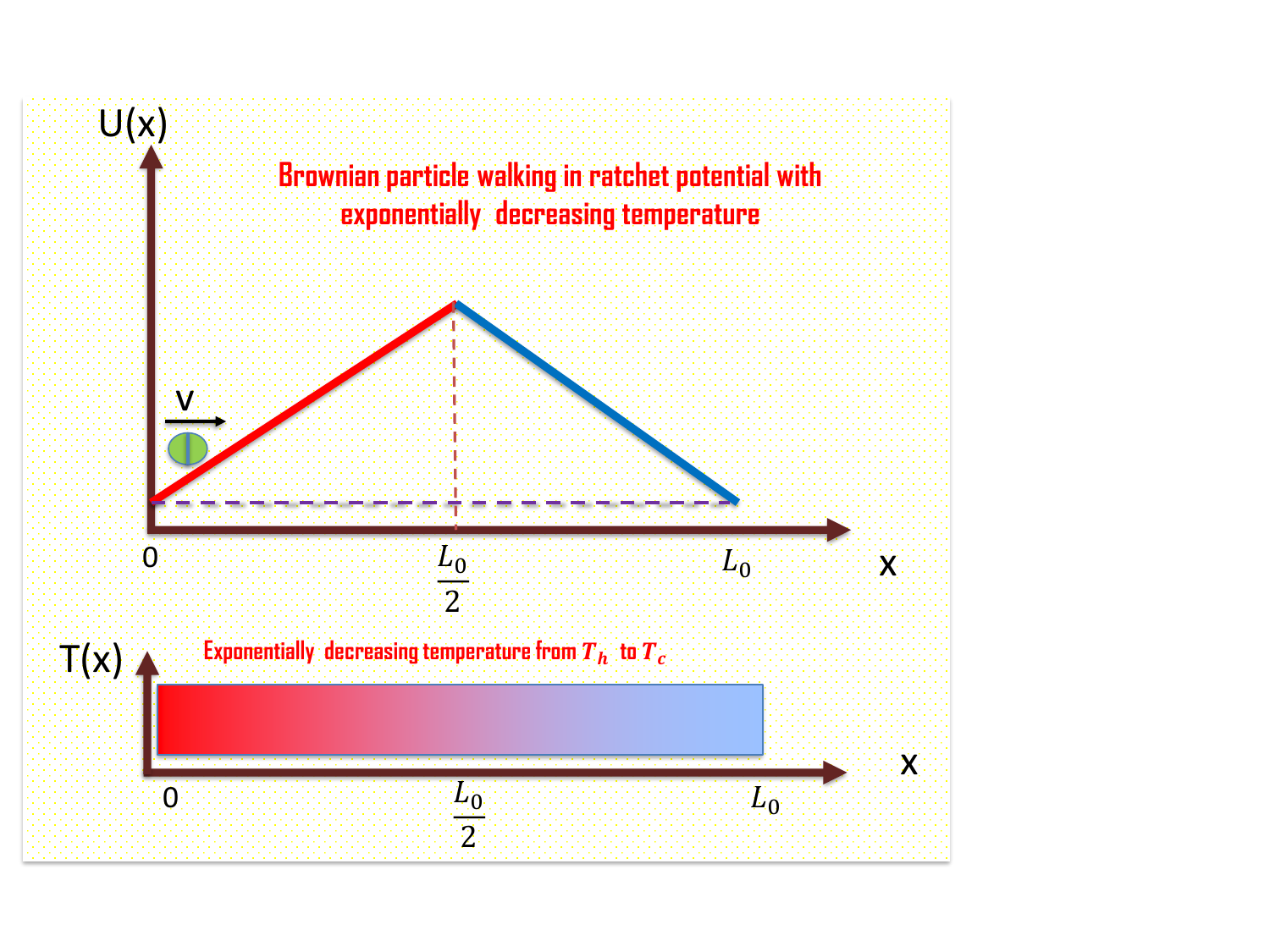}}
\caption{(Color online)  Schematic diagram for a Brownian particle in a piecewise linear potential in the absence of external load.  The temperature  decreases  exponentially from $T_h$ to $T_c$.
 } 
\end{figure}

The dynamics of  the Brownian particles are governed by the Smoluchowski equation:
\begin{equation}
\frac{\partial P(x, t)}{\partial t} = \frac{\partial}{\partial x} \left[\frac{1}{\gamma} \left(U'(x)P(x, t) + \frac{\partial T(x)P(x, t)}{\partial x}\right)\right],
\end{equation}
where $P(x, t)$ is the probability density, $U'(x) = \frac{dU(x)}{dx}$, and $\gamma$ denotes  viscous friction. At steady state
\begin{equation}
J=-\frac{1}{\gamma} \left[U'(x)P^s(x) + \frac{d}{dx}(T(x)P^s(x))\right]
\end{equation}
where $U(x)=U_s(x)+fx$.

\section{Steady state Current, Efficiency, and Coefficient of Performance of Refrigerator}

It is important to note that in the absence of symmetry-breaking fields, no net flow of particles can be  obtained. The unidirectional motion of a particle is attainable only in the presence of externally acting loads or inhomogeneous temperature distributions. Hereafter, whenever we plot figures, we use  the dimensionless rescaled parameters temperature $\tau = T_h/T_c$, barrier height $\bar{U}_0 = U_0/T_c$,  load $\lambda = fL_0/T_c$ and length $\bar{x} = x/L_0$. For simplicity, the bars are omitted. Moreover, in this study, the viscous friction coefficient $\gamma$ and Boltzmann constant $k_B $, are both set to unity.  The general expression  for the steady-state current $J$   in any periodic potential with or without a load is  reported in \cite{c9,c11}.  Following the same approach, we find  the steady state current J  as 
\begin{equation}   
J= {-F\over G_{1}G_{2}+HF}.
\end{equation} 
Detailed derivations and expressions are provided in Appendix I.

 The velocity $V$ is then expressed as
$
V = L_0J,
$
where $J$ denotes the steady state current. 
We also compare the analytical results with the simulation outcomes for both the short- and long-time cases. The dynamics of the system are analyzed by integrating the Langevin equation and employing a Brownian dynamics simulation. In the simulation, a Brownian particle is initially positioned within one of the potential wells. The trajectories of the particles are then simulated for different time steps $\Delta t$ and total time length $t_{\text{max}}$. To ensure numerical accuracy, up to $1 \times 10^6$ ensemble averages have been obtained.

\begin{figure}[ht]
\centering

{

    \includegraphics[width=6cm]{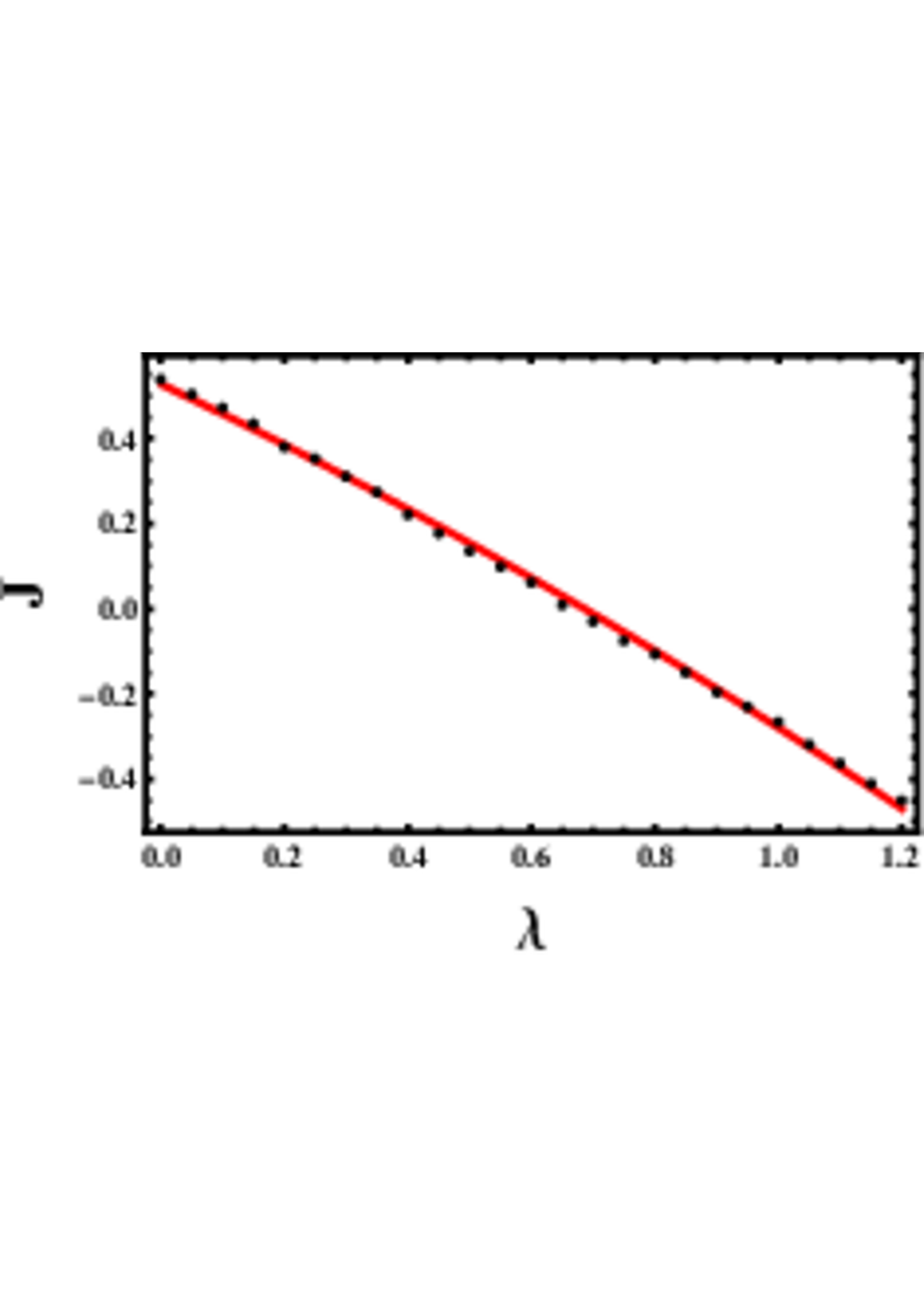}
}

\caption{ (Color online) The plot illustrates the current $J$ for the exponential thermal arrangement as a function of $\lambda$  for fixed parameters $\tau = 2.0$ and $U_0 = 2.0$. The solid line represents the plot derived from the analytical result, whereas the dotted line indicates the plot obtained from the Brownian dynamics simulation at steady state.
} 
\label{fig:sub} 
\end{figure}

\begin{figure}[ht]
\centering
\includegraphics[width=0.32\textwidth]{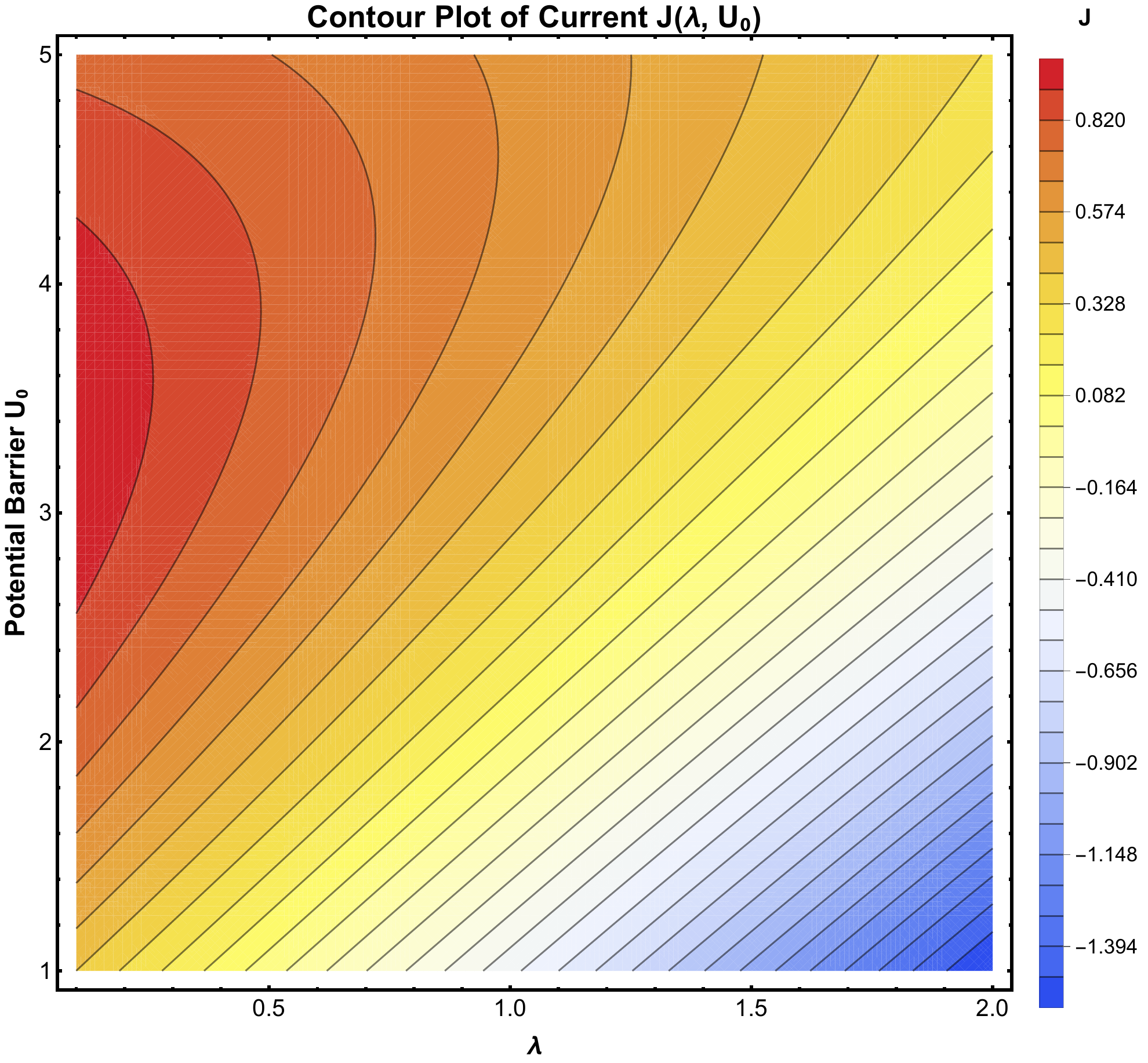}\hspace{0.01\textwidth}
\includegraphics[width=0.32\textwidth]{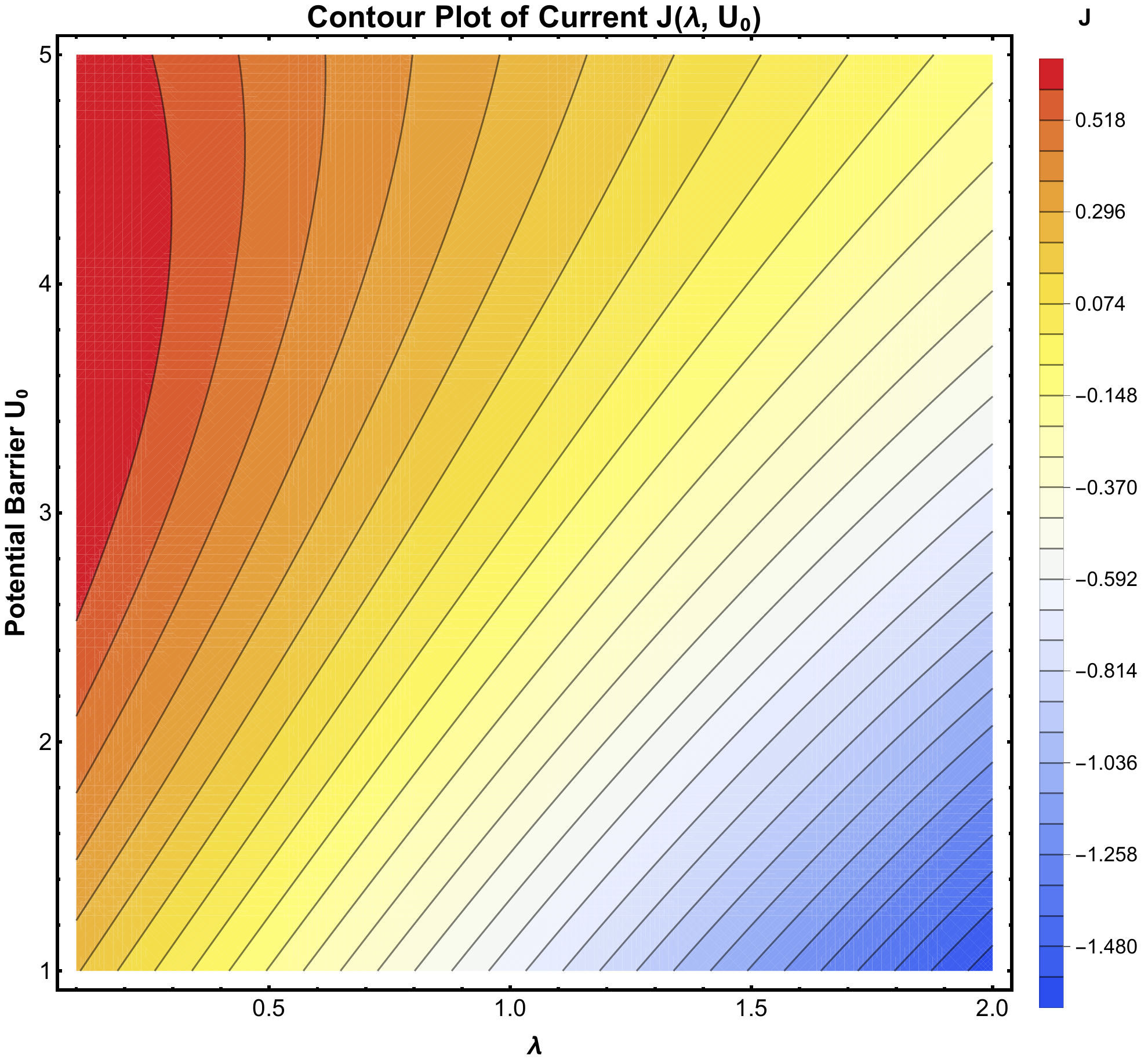}\hspace{0.01\textwidth}
\includegraphics[width=0.32\textwidth]{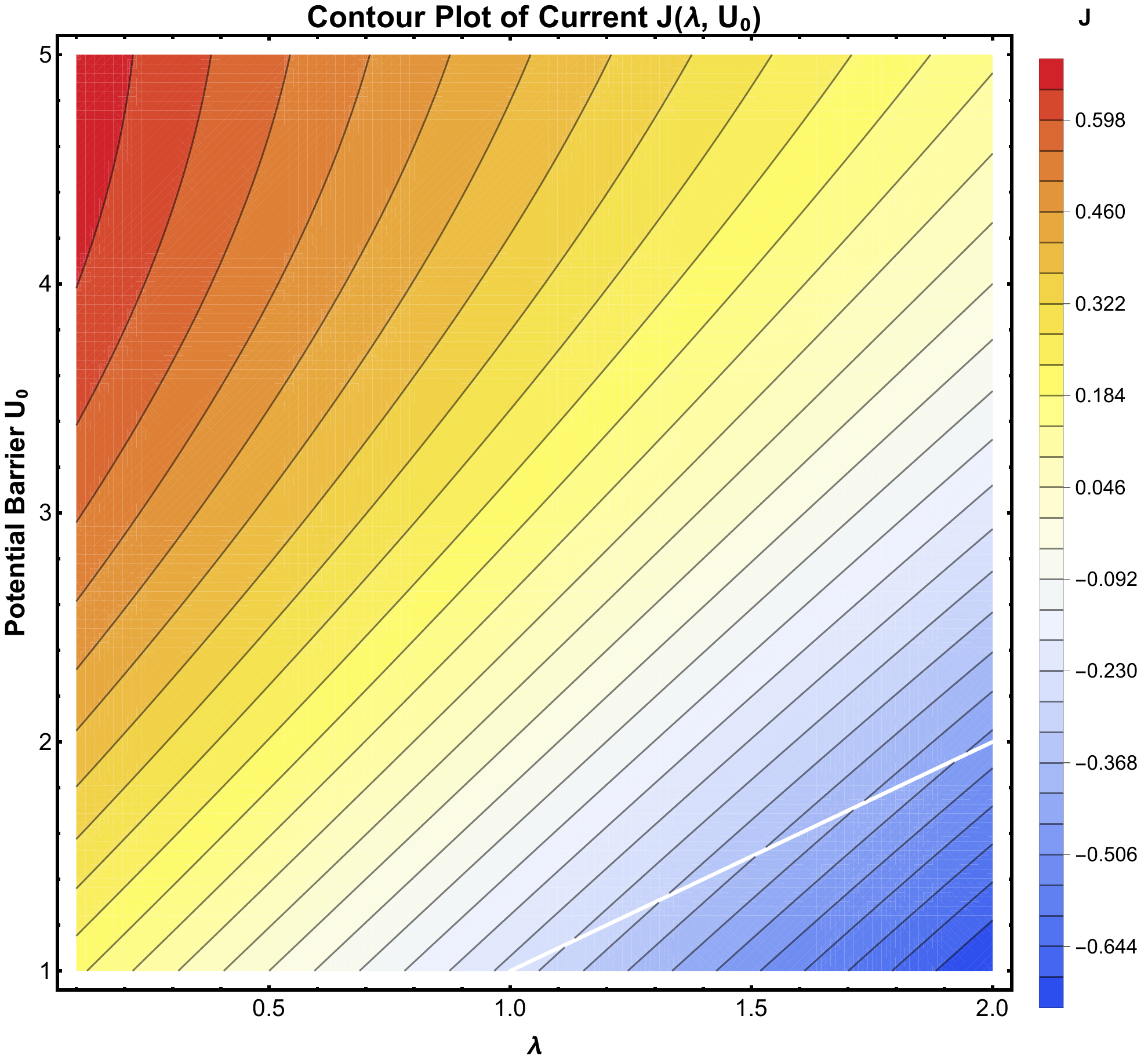}
\vspace{1em}
\caption{
Contour plots of the steady-state current $J$ as a function of external force $\lambda$ and potential barrier height $U_0$ for the three thermal configurations at a fixed rescaled temperature $\tau = 2$. (top) Exponentially decreasing temperature profile, (middle) linearly decreasing profile, and (bottom) piecewise constant profile. Color indicates current direction and strength; the zero-current contour ($J = 0$) marks the stall line between heat engine and refrigerator operation.}
\label{fig:J_all_profiles}
\end{figure}

 In Figure 2, we plot the results for both the analytical and simulation cases  at  a steady state. The figure shows that the two results are in agreement.   Figure~\ref{fig:J_all_profiles} presents contour plots of the steady-state current $J$ as a function of the external force $\lambda$ and potential barrier height $U_0$, evaluated at fixed   $\tau = 2$, for three distinct thermal configurations: exponentially decreasing (top), linearly decreasing (center), and piecewise constant (bottom) temperature profiles. In all cases, the magnitude and direction of the current exhibit a pronounced dependence on both $\lambda$ and $U_0$. As the barrier height increases, the current initially rises due to enhanced rectification and thermally assisted barrier crossing, reaches a maximum at intermediate values of $U_0$, and subsequently declines as the potential barrier becomes too large to support efficient transport. Among the three configurations, the exponential temperature profile yields the highest current values across a broad region of the parameter space, reflecting its enhanced transport efficiency enabled by a smooth and spatially coherent thermal gradient. The linearly decreasing profile produces moderately strong currents, whereas the piecewise constant profile results in a substantially reduced current magnitude. This contrast highlights the influence of thermal continuity: smooth temperature variations promote sustained asymmetric energy transfer, while discontinuities suppress directional coherence. In each contour plot, a clearly defined zero-current line ($J = 0$) separates the operational regimes of the system, distinguishing the heat engine behavior from that of refrigeration. At low values of $\lambda$, the current is positive and decreases with increasing load, which is characteristic of the engine operation. At higher loads, current reversal occurs, signifying a transition to refrigerator behavior. These results underscore the critical role of both the thermal gradient and energy landscape in governing nonequilibrium transport and thermodynamic functionality in Brownian heat engines.

  Similar to the rate of entropy production, the degree of irreversibility can be measured using the efficiency of a Brownian motor. To compare the efficiency among the three different thermal arrangements, let us first write the efficiency of the motor as 
\begin{equation}
\eta = \frac{W^s}{Q_{\text{in}}},
\end{equation}
where the work done is given by
\begin{equation}
W^s = fL_0,
\end{equation}
while, to surmount the ratchet potential, the particle should receive an input energy
\begin{equation}
Q_{\text{in}} = U_0 + \frac{fL_0}{2},
\end{equation}
from the left-hand side of the potential.    

Before determining the efficiency at the quasistatic limit, let us determine the stall force, where the current ($J$) equals zero, for each of the two cases. After some algebra, the stall force for the exponential thermal arrangement is given by
\begin{equation}
f' = \frac{2 (-1 + \sqrt{T_h / T_c}) U_0}{L (1 + \sqrt{T_h / T_c})}.
\end{equation}
The stall force for the piecewise constant thermal arrangement is given as
\begin{equation}
f' = -\frac{(T_c - T_h) U_0}{L (T_c + T_h)}.
\end{equation}

Our algebraic analysis reveals that the efficiency is significantly low for temperature that decreases exponentially. This observation is further elucidated by calculating the efficiency in the quasistatic limit. For systems where the temperature decreases exponentially, in the quasistatic limit (substituting Eq. (11) into Eq. (8)), the efficiency is given by
\begin{equation}
\eta = 1 - \frac{1}{\sqrt{T_h / T_c}},
\end{equation}
where $\eta$ is exactly equal to the efficiency of an endoreversible heat engine. The endoreversible efficiency assumes that irreversibilities are confined to heat transfer processes, leading to better performance than that in real-world engines, yet always remaining below the Carnot efficiency.

As anticipated, for a piecewise constant thermal arrangement at the quasistatic limit (substituting Eq. (12) into Eq. (8)), the efficiency asymptotically approaches the Carnot efficiency.
\begin{equation}
\eta = 1 - \frac{T_c}{T_h}.
\end{equation}

\begin{figure}[ht]
\centering
{
    \includegraphics[width=8cm]{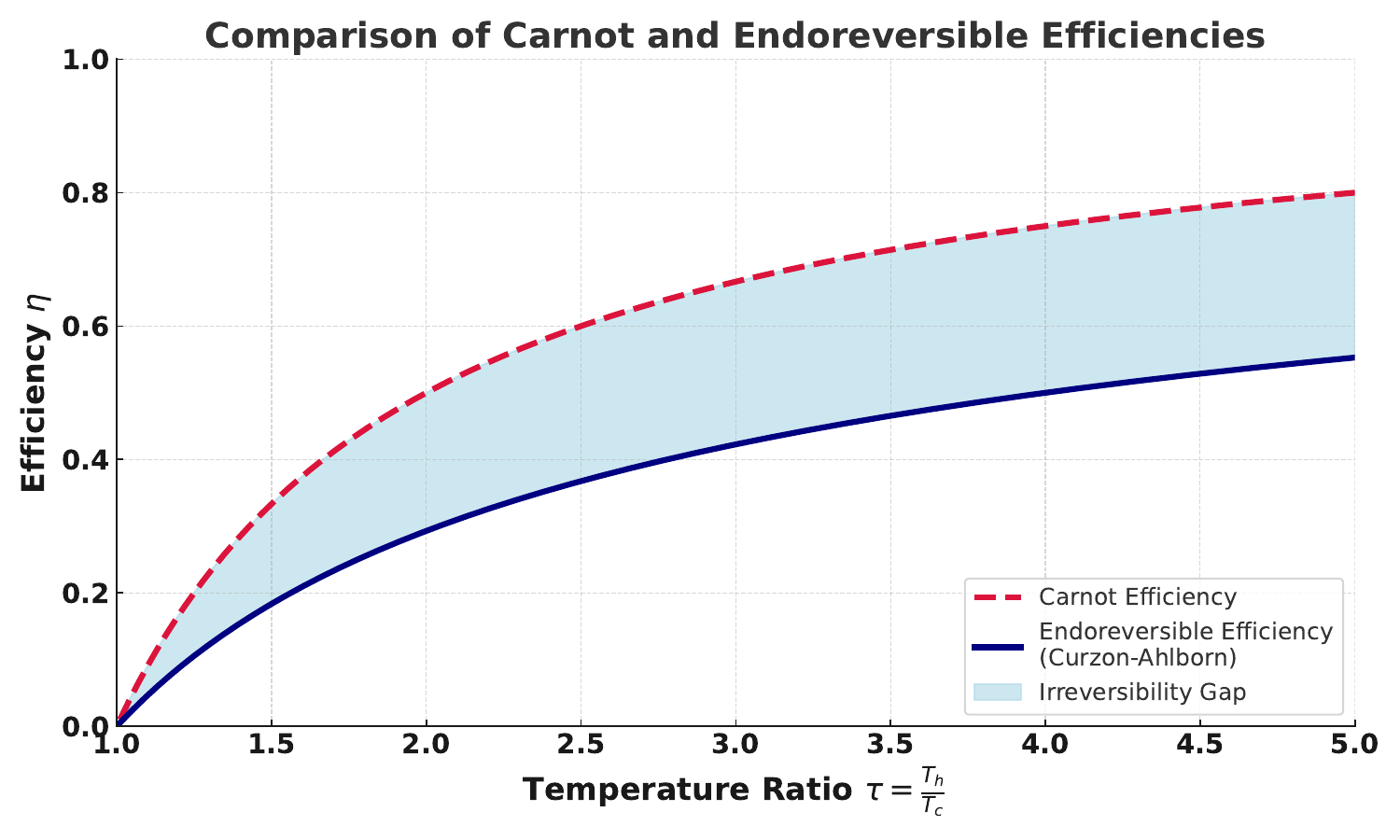}
}
\caption{ (Color online) The plot presents  the endoreversible efficiency (blue line) and the Carnot efficiency (red dashed line) as functions of rescaled temperature  $\tau=T_h/T_c$.
} 
\label{fig:sub} 
\end{figure}

Moreover whenever the engine functions as a refrigerator, the coefficient of performance (COP)  can be written as  
\begin{equation}
COP = \frac{Q_c}{W^S},
\end{equation}
where
\begin{equation}
Q_{c} = U_0 - \frac{fL_0}{2}.
\end{equation}
where $Q_c$ denotes the amount of heat transferred to the coldest part of the bath. Our algebraic analysis reveals that the COP is significantly lower for a temperature that decreases exponentially. This observation can be further clarified by calculating the COP at the quasistatic limit, as follows: For systems with exponentially decreasing temperature, in the quasistatic limit (substituting Eq. (11) into Eq. (15)), the COP is given by
\begin{equation}
COP = \frac{1}{\sqrt{\frac{T_h}{T_c} - 1}}.
\end{equation}
This expression is equivalent to the coefficient of performance of an endoreversible heat engine.

Furthermore, in the case of a piecewise constant thermal arrangement, where the system is coupled with both hot and cold baths at the quasistatic limit (substituting Eq. (12) into Eq. (15)), the typical Carnot refrigerator is obtained as follows
\begin{equation}
COP = \frac{T_c}{T_h - T_c}.
\end{equation}

In this study, we analyze the thermodynamic performance of a single Brownian particle under an exponentially decreasing temperature profile. Despite the fundamental differences from classical systems, we find that the efficiency and Coefficient of Performance (COP) of our model converge to those of an endoreversible engine.  In this regime,   because the particle  moves  very slowly, it remains near equilibrium. Consequently,  the efficiency aligns with the endoreversible efficiency, whereas the COP approaches that of an endoreversible refrigerator. These results highlight the universal applicability of classical thermodynamic principles even in microscopic stochastic systems.   

Next,   in Fig. 4, we plot the dependence of  the endoreversible efficiency  and Carnot efficiency on $T_h$ and $T_c$. The figure shows that the Carnot efficiency (red dashed line)  is considerably larger than the endoreversible efficiency (blue line).

In this section, we  explore  the thermodynamic features of  a Brownian heat engine  that  moves in a periodic ratchet potential coupled  with an exponentially decreasing temperature gradient. At the quasistatic limit, the system approaches the efficiency of an endoreversible heat engine.  Beyond providing theoretical insights, this study has important implications for microscale and nanoscale transport applications, particularly in energy harvesting and molecular-motor design. A continuous thermal gradient facilitates directed motion even in the  absence of external forces   that mimic   biological transport mechanisms and artificial systems. We believe that these findings provide a framework for optimizing transport in microfluidic devices, nanoscale sensors and Brownian ratchets.

\section{Entropy, Entropy Production Rate and Free Energy}

In this section, we explore the dependence of key thermodynamic  relations, such as entropy, entropy production, and entropy production  rates, on the system parameters.  The non-equilibrium Gibbs entropy 
\begin{equation}
S(t) = -\int P_s(x) \ln P_s(x) \, dx
\end{equation}
is a fundamental concept in statistical mechanics that generalizes the entropy definition for systems that are out of equilibrium. 
\begin{figure}[ht]
\centering
{
    \includegraphics[width=10cm]{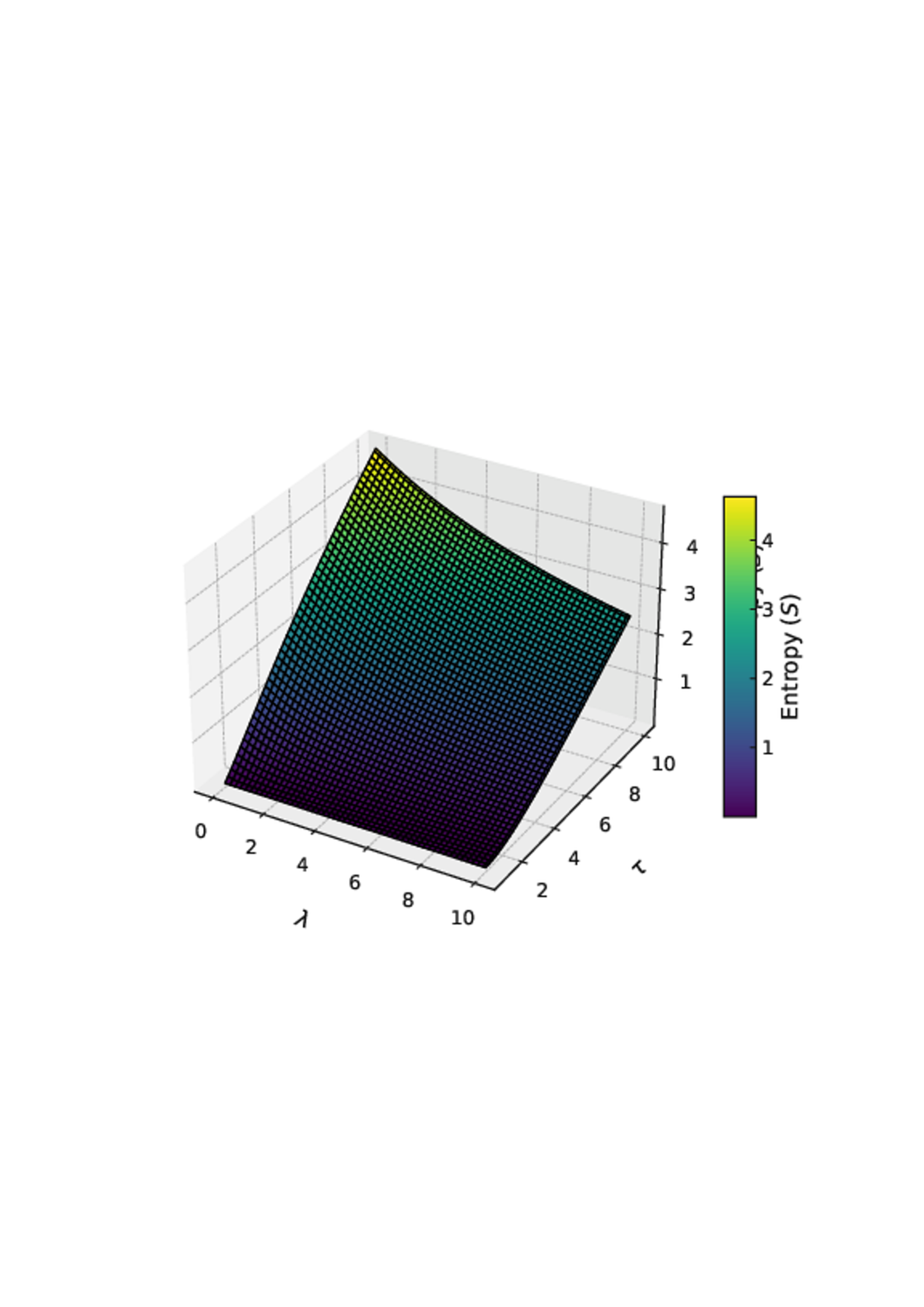}
}
\caption{ (Color online) Entropy ($S$) as a function of load ($\lambda$) and rescaled temperature  ($\tau$) for fixed values of  $U_0 = 2$. The 3D plot shows that the  entropy increases as the load decreases and when  rescaled temperature increases. } 
\label{fig:sub} 
\end{figure}
The corresponding probability distribution $P_s(x)$  is derived and presented in Appendix II.

In Fig. 5, we plot the entropy $S$ as a function of the load  and  rescaled  temperature ($\tau$) for the exponentially decreasing temperature case. The figure shows that entropy  decreases  as the load steps up and  as temperature increases. Moreover, our analysis indicates that the entropy for the exponentially decreasing temperature case is considerably larger than that for the piecewise constant-temperature case. Appendix 2 presents the derivation of the probability distribution for the exponentially decreasing case. 

The time evolution of the non-equilibrium Gibbs entropy follows an entropy balance equation that captures the competition between entropy production  and entropy dissipation rates in a system. The entropy change is given by  \cite{c20}
\begin{eqnarray}
\frac{dS(t)}{dt} &=& \dot{e}_{p} - \dot{h}_{d} \\
&=& \int \left( \frac{J^{2}}{P_s(x)T(x)} + J \frac{U'(x)}{T(x)} + J \frac{T'(x)}{2T(x)} \right) dx,
\end{eqnarray}
where the entropy production rate $\dot{e}_{p}$ and dissipation rate $\dot{h}_{d}$ are defined as
\begin{eqnarray}
\dot{e}_{p} &=& \int \frac{J^{2}}{P_s(x)T(x)} dx,
\end{eqnarray}
and 
\begin{eqnarray}
\dot{h}_{d} &=& \int \left( J \frac{U'(x)}{T(x)} + J \frac{T'(x)}{2T(x)} \right) dx.
\end{eqnarray}
Unlike the isothermal scenario, the term $J \frac{T'(x)}{2T(x)}$ introduces an additional contribution. In the steady state, where $\frac{dS(t)}{dt} = 0$, it follows that $\dot{e}_{p} = \dot{h}_{d} > 0$. In the stationary state, approaching equilibrium, $J = 0$ ensures a detailed balance condition, yielding $\dot{e}_{p} = \dot{h}_{d} = 0$. Moreover, at the quasistatic limit for the exponentially decreasing temperature case, where 
$f \to f' = \frac{2(-1 + T_h/T_c)U_0}{L(1 + \sqrt{T_h/T_c})}$, 
or for the piecewise constant temperature case, where 
$f \to f' = \frac{-(T_c - T_h)U_0}{L(T_c + T_h)}$,  
The entropy production rate and entropy extraction rate approach zero. The exact expressions for $J(x,t)$ and $P_s(x)$, although lengthy, are provided in Appendices I and II, respectively.

To appreciate this result, let us analyze the entropy production rate, or equivalently the entropy extraction rate, for the isothermal case, where $T_h \to T_c$. For the isothermal case, for both exponentially decreasing temperature and piecewise constant temperature configurations, we get
the entropy production  or extraction rates as
\begin{widetext}
\begin{equation}
\dot{e}_p = \dot{h}_d = \frac{f \left(-f^2 L^2 + U_0^2\right)^2 \sinh\left(\frac{f L}{T_c}\right)}{2 L T_c \left[2 T_c U_0^2 \left(-\cosh\left(\frac{f L}{T_c}\right) + \cosh\left(\frac{U_0}{T_c}\right)\right) + f L (f L - U_0)(f L + U_0) \sinh\left(\frac{f L}{T_c}\right)\right]}.
\end{equation}
\end{widetext}
The expression  $\dot{e}_p = \dot{h}_d>0  $. 
In the limit $U_0 \to 0$, the above equation simplifies to
\begin{equation}
\dot{e}_p = \frac{f^2}{2 T_c}.
\end{equation}
This indicates that the entropy production rate is directly proportional to the square of the external force $f$ and inversely proportional to the cold bath temperature $T_c$.
In the limit $f \to 0$, the above expression  approaches 
\begin{equation}
\dot{e}_p = 0.
\end{equation}
This indicates  that in  the absence of an external force,    entropy production  vanishes.  This behavior aligns with the fundamental thermodynamic expectation that a system operates reversibly in the absence of external forces with no net entropy production. 

\begin{figure}[ht]
\centering
\includegraphics[width=8cm]{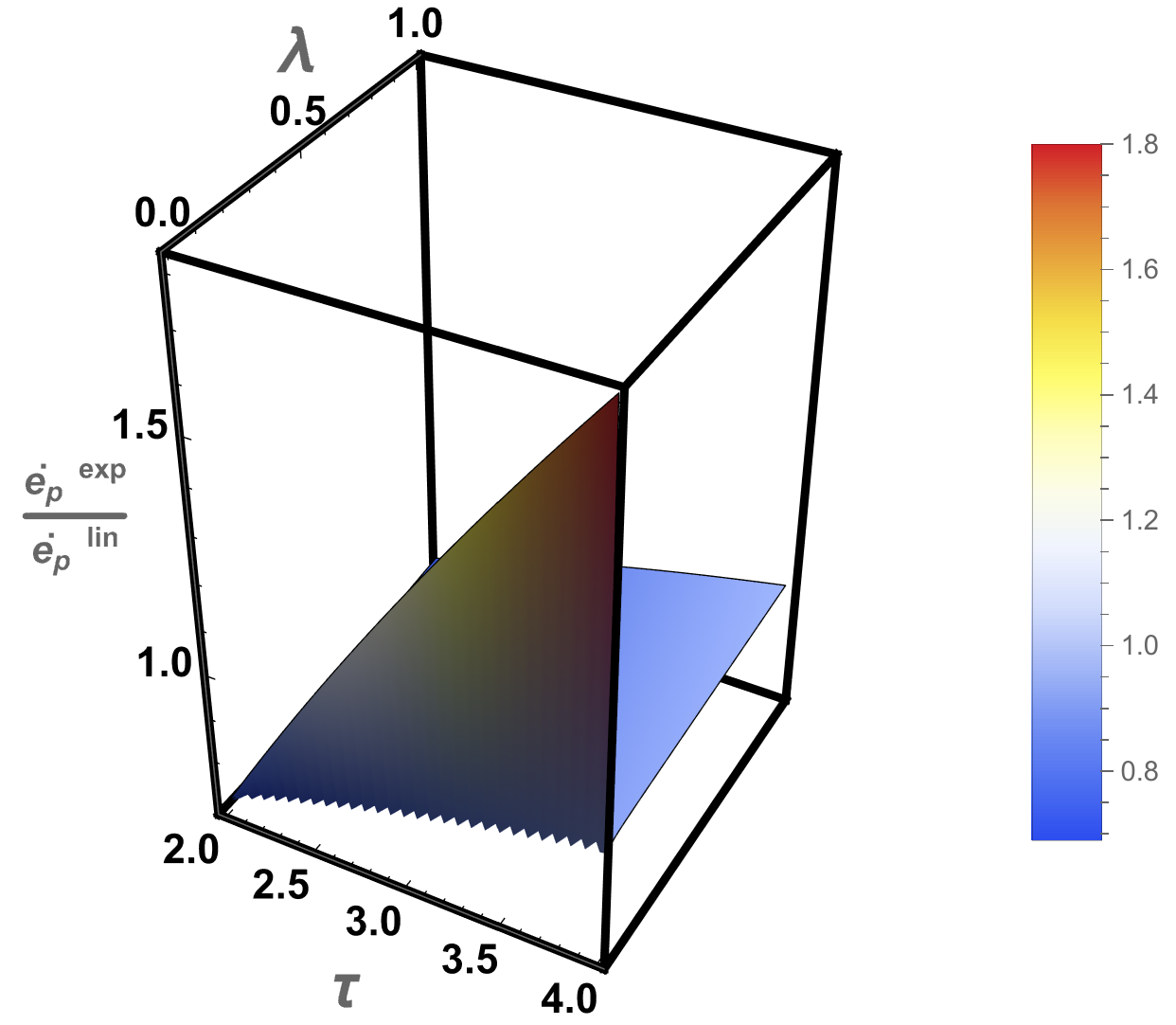}
\caption{
The plot of the ratio of entropy production rates, $\dot{e}_p^{\mathrm{(exp)}} / \dot{e}_p^{\mathrm{(lin)}}$, as a function of the rescaled  temperature $\tau$ and external force $\lambda$. The plot shows that the exponential temperature profile results in a consistently higher entropy production rate across much of the parameter space, indicating enhanced thermodynamic activity and stronger transport under a smoother thermal gradient. Regions of elevated ratios reflect the pronounced irreversibility associated with faster dynamics in the exponential case, in contrast to the more moderate dissipation in the linearly decreasing profile.}
\label{fig:entropy_ratio}
\end{figure}
Figure~6 presents the  plot of the ratio of the entropy production rates, $\dot{e}_p^{\mathrm{(exp)}} / \dot{e}_p^{\mathrm{(lin)}}$, as a function of the rescaled temperature, $\tau$ and external load, $\lambda$, with the potential barrier fixed at $U_0 = 4.0$. The plot clearly shows that the exponential thermal profile yields a higher entropy production rate across a broad region of the parameter space than the linearly decreasing temperature profile. This enhancement reflects the stronger thermodynamic activity and faster particle transport enabled by the smooth and spatially coherent gradient in the exponential case. The ratio increases, particularly in regions of high $\tau$ and moderate-to-low load, where the exponential profile maintains a more effective temperature bias. These results highlight the impact of the thermal profile shape on the irreversibility and performance of Brownian heat engines.

We now  study the model system further by exploring  $\dot{E}_{p}(t)$ and $\dot{H}_{d}(t)$.   The term related  to   the heat dissipation rate is given as 
\begin{eqnarray}
\dot{H}_{d} &=& \int \left( J U'(x) + \frac{J T'(x)}{2} \right) dx.
\end{eqnarray}
The term $\dot{E}_{p}$, associated with $\dot{e}_{p}$  and it can be written as 
\begin{eqnarray}
\dot{E}_{p} &=& \int \frac{J^{2}}{P_s(x)} dx.
\end{eqnarray}
The entropy balance equation becomes
\begin{eqnarray}
\frac{dS^{T}(t)}{dt} &=& \dot{E}_{p} - \dot{H}_{d} \\
&=& \int \left( \frac{J^{2}}{P_s(x)} + J U'(x) + J T'(x) \right) dx.
\end{eqnarray}
Our analysis also indicates that the entropy production rate $\dot{E}_p=\dot{H}_d$ is consistently higher in the exponential thermal arrangement than in the linear case.

Moreover, the rate of  internal energy is defined as
\begin{eqnarray}
\dot{E}_{\text{in}} = \int J U'_{s}(x) dx.
\end{eqnarray}
For a Brownian particle that operates  under spatially varying temperature, the total work done is given by
\begin{eqnarray}
\dot{W} &=& \int \left( J f + \frac{J T'(x)}{2} \right) dx.
\end{eqnarray}
The first law of thermodynamics is elegantly written as 
\begin{eqnarray}
\dot{E}_{\text{in}} = -\dot{H}_{d}(t) - \dot{W}.
\end{eqnarray}
Consequently, the change in internal energy is given as 
\begin{equation}
\Delta E_{\text{in}} = -\int_{0}^{t} \left( \dot{H}_{d}(t) + \dot{W} \right) dt.
\end{equation}

As elaborated in \cite{c20,c21,c22}, the free-energy rate is given by $\dot{F} = \dot{E} - T \dot{S}$ for the isothermal case and $\dot{F} = \dot{E} - \dot{S}^{T}$ for the non-isothermal case, where $\dot{S}^{T} = \dot{E}_{p} - \dot{H}_{d}$.  We then write the free energy dissipation rate  as  
\begin{eqnarray}
\dot{F} &=& \dot{E}_{\text{in}} - \dot{S}^{T} \\
&=& \dot{E}_{\text{in}} - \dot{E}_{p} + \dot{H}_{d},
\end{eqnarray}
after some algebra,  the change in free energy  has a form 
\begin{eqnarray}
\Delta F(t) &=& -\int_{0}^{t} \left( \dot{W} + \dot{E}_{p}(t) \right) dt.
\end{eqnarray}

The main result of our study shows  that entropy production and free energy dissipation are considerably large for  an exponentially decreasing temperature gradient  compared with those in a piecewise constant temperature profile. We show  that both the entropy production and extraction rates increase with the applied load and  temperature. Our results indicate  that at a steady state, the entropy production and dissipation rates  balance each other. The exponentially decreasing temperature profile also  results in a significantly higher entropy production rate compared to the piecewise constant case. 
Additionally, we find that the entropy production rate is considerably higher in an exponentially decreasing thermal arrangement, which enhances  the mobility of the particle but also leads to greater irreversibility. This suggests that an exponential thermal profile can drive non-equilibrium processes more effectively with a higher velocity but  with less efficient energy utilization compared to a piecewise constant temperature profile.  We believe that understanding these effects is particularly relevant for optimizing microscale thermal devices, energy harvesting systems, and nanoscale transport mechanisms because managing entropy production and irreversibility is crucial for achieving higher performance and efficiency.

\section{Heuristic Treatment of Entropy and Current for Repetitive Networks Sharing Both Ends}

In this work, extending our previous study \cite{c22}, we further analyze the thermodynamic features of a single Brownian particle moving along $M$ Brownian ratchets arranged in a complex network. Each ratchet potential is either coupled with hot and cold reservoirs or a heat bath, where its temperature decreases exponentially along the reaction coordinate.
The analytical results reveal that the rates of thermodynamic quantities, such as the velocity $V$, entropy production $\dot{e}_p(t)$, and entropy extraction $\dot{h}_d(t)$ are independent of the network size at steady state, as reconfirmed by the complex generating functions \cite{c22}.
On the contrary, the thermodynamic relations, including entropy $S$, entropy production $e_p(t)$, and entropy extraction $h_d(t)$ of the system, increase with the network size $M$, even at a steady state.

Consider  Brownian particles operating in networks with $N$ ratchet potentials, under an exponentially decreasing temperature profile.   $N$  networks that  are repetitive  shear the same endpoints. The global probability density for a network with $N$ branches sharing both ends is 
\begin{equation}
P_{\text{total}}(x) = \sum_{k=1}^N P_k(x),
\end{equation}
where $P_k(x)$ denotes the probability density of the $k$ th branch. The total probability is normalized as
\begin{equation}
\int_0^L P_{\text{total}}(x) \, dx = 1.
\end{equation}
Substituting $P_{\text{total}}(x)$ into the normalization condition yields
\begin{equation}
\int_0^L \sum_{k=1}^N P_k(x) \, dx = 1.
\end{equation}
The normalized global probability density is
\begin{equation}
P(x) = \frac{\sum_{k=1}^N P_k(x)}{\int_0^L \sum_{k=1}^N P_k(x) \ dx},
\end{equation}
On the other hand, 
the entropy of the system is defined as
\begin{equation}
S = -\int_0^L P_{\text{total}}(x) \ln P_{\text{total}}(x) \, dx.
\end{equation}
Substituting $P_{\text{total}}(x) = N  P_k(x)$ gives
\begin{equation}
S = -\int_0^L \left( N \cdot P_k(x) \right) \ln \left( N  P_k(x)\right) \, dx.
\end{equation}
Expanding the logarithm yields
\begin{equation}
\ln \left( N  P_k(x) \right) = \ln N + \ln P_k(x),
\end{equation}
and substituting yields
\begin{equation}
S = -N \ln N \int_0^L P_k(x) \, dx - N \int_0^L P_k(x) \ln P_k(x) \, dx.
\end{equation}
Using the normalization condition $\int_0^L P_k(x) \, dx = \frac{1}{N}$, the first term simplifies to
$
-N \ln N \cdot \frac{1}{N} = -\ln N.
$
Thus, the entropy becomes
\begin{equation}
S = -\ln N - N \int_0^L P_k(x)\ln P_k(x) \, dx.
\end{equation}
As one can see, the entropy production depends on the network size $N$.

Furthermore, the current in the $k$-th branch is given by
\begin{equation}
J_k = A_k(x) P_k(x) - \frac{\partial}{\partial x} \left[ D_k(x) P_k(x) \right],
\end{equation}
where  $
A_k(x) = -\frac{\partial U_k(x)}{\partial x} + f$  and  $
D_k(x)$  denotes  the diffusion coefficient.

The total current across all branches is
\begin{equation}
J_{\text{total}} = \sum_{k=1}^N J_k.
\end{equation}
For symmetric branches sharing flux equally
\begin{equation}
J_k = \frac{J_{\text{total}}}{N}.
\end{equation}
Substituting back verifies
\begin{equation}
J_{\text{total}} = J_k.
\end{equation}
This indicates that  the particle current, or  equivalently, the velocity, is independent of the network size $N$.  

In order to check the dependence of   the entropy production rate on the network size,   let us write the entropy  for each branch as 
\begin{equation}
\dot{S}_k = \int_0^L \frac{J_k^2}{D(x) P_k(x)} \, dx,
\end{equation}
where $J_k$ is the local current, $P_k(x)$ is the probability density, and $D(x)$ is the diffusion coefficient. Summing over $N$ branches yields
\begin{equation}
\dot{S}_{\text{total}} = \sum_{k=1}^N \dot{S}_k = \sum_{k=1}^N \int_0^L \frac{J_k^2}{D(x) P_k(x)} \, dx.
\end{equation}
Substituting $J_k = \frac{J_{\text{total}}}{N}$ and $P_k(x) = \frac{P_{\text{total}}(x)}{N}$ gives
\begin{equation}
\dot{S}_{\text{total}} = \int_0^L \frac{J_{\text{total}}^2}{D(x) P_{\text{total}}(x)} \, dx.
\end{equation}
Thus, the entropy production rate depends on $J_{\text{total}}$, $D(x)$, and $P_{\text{total}}(x)$ but it is  independent of $N$. This clearly indicates that the entropy production rate  does not depend on the network size.  

Similarly, 
the entropy extraction rate is defined as
\begin{equation}
\dot{h}_{d} = \int_0^L \left( J(x) \frac{U'(x)}{T(x)} + J(x) \frac{T'(x)}{2T(x)} \right) dx,
\end{equation}
where $J(x)$ is the local current, $U'(x)$ is the potential gradient, and  $T(x)$ is the temperature.   \( T'(x) \) denotes the derivative of the temperature with respect to position, i.e., \( T'(x) = \frac{dT}{dx} \). Substituting $J(x) = \frac{J_{\text{total}}}{N}$
\begin{equation}
\dot{h}_{d} = \frac{J_{\text{total}}}{N} \int_0^L \left( \frac{U'(x)}{T(x)} + \frac{T'(x)}{2T(x)} \right) dx.
\end{equation}
Summing over $N$ branches
\begin{equation}
\dot{h}_{d, \text{total}} = \sum_{k=1}^N \dot{h}_{d}.
\end{equation}
The $N$ factors cancel, leaving
\begin{equation}
\dot{h}_{d, \text{total}} = J_{\text{total}} \int_0^L \left( \frac{U'(x)}{T(x)} + \frac{T'(x)}{2T(x)} \right) dx.
\end{equation}
Thus, the entropy extraction rate is independent of $N$ and depends only  on $J_{\text{total}}$, $U'(x)$, and $T(x)$.

\section{ Brownian  Dynamic Simulation of Short-Time Behavior}
The short-time dynamics of the system are analyzed  via  Brownian dynamics simulations. In these simulations, a Brownian particle is initially positioned in one of the potential wells. The particle's trajectories are computed by varying the time steps $t$ and the total simulation time $t_{\text{max}}$. To ensure numerical accuracy, up to $1 \times 10^6$ ensemble averages have been obtained.
\begin{figure}[ht]
\centering
   { \includegraphics[width=6cm]{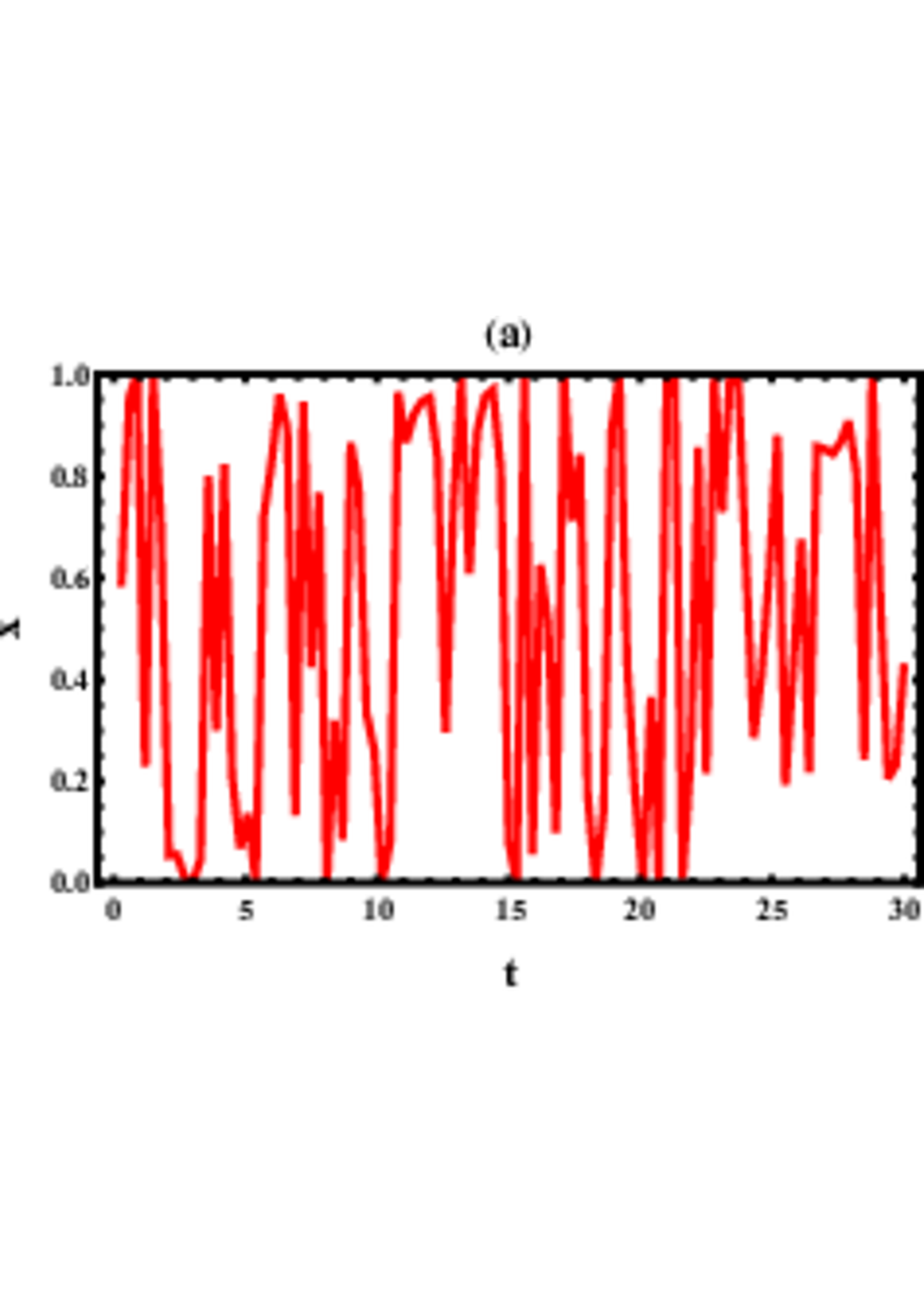}}
\hspace{1cm}
{
    \includegraphics[width=6cm]{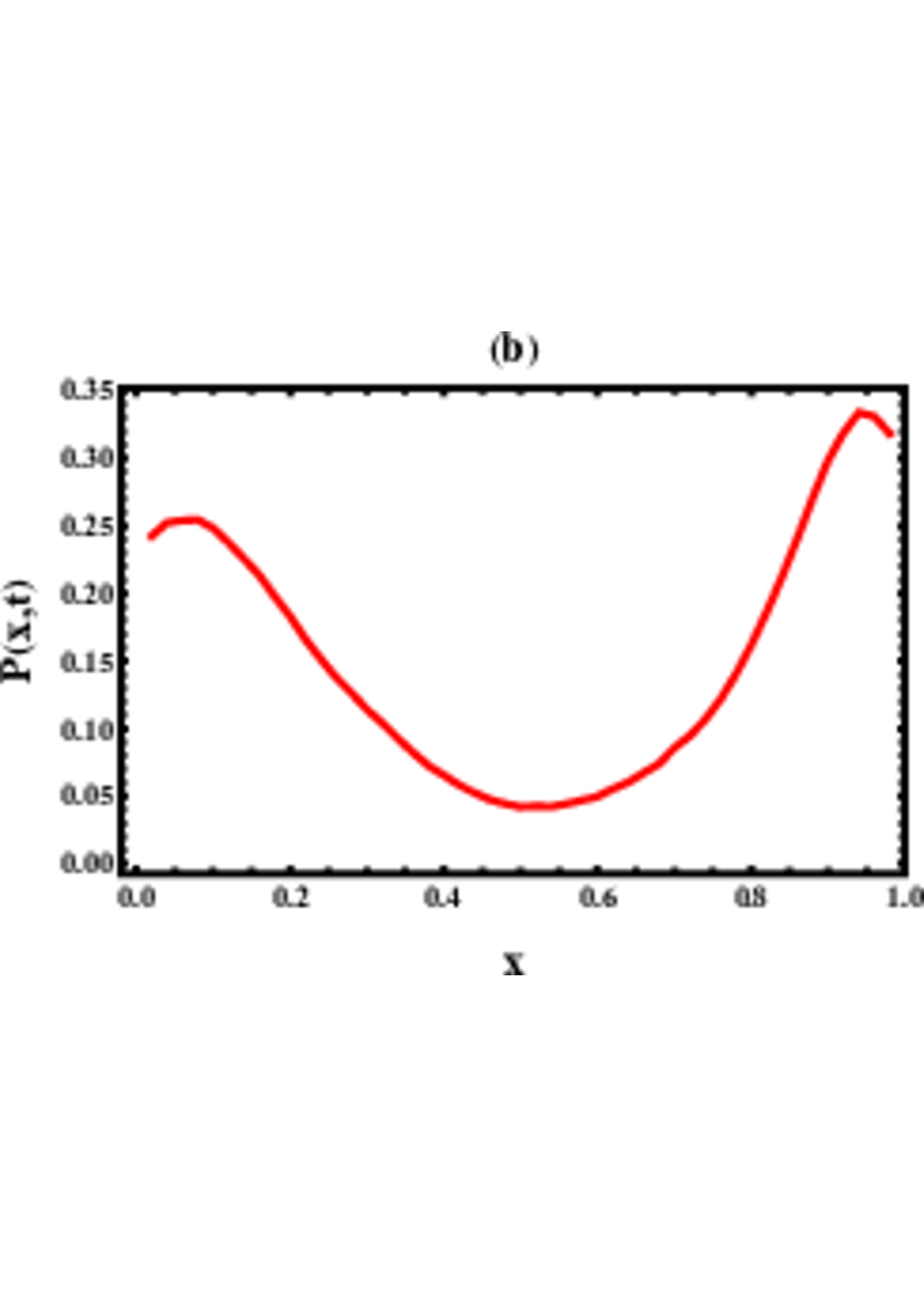}
}
\caption{ (Color online) (a) The trajectory $x(t)$ as a function of time $t$ is evaluated through numerical simulations for a fixed external force $f = 0.0$, $U_0 = 4.0$, and $\tau = 1.0$. We fix  the other  parameters as   $U_0 = 4.0$ and $\tau = 1.0$. (b) The probability distribution $P(x, t)$ as a function of time $t$ is analyzed for fixed  values of  $f = 0.0$, $U_0 = 4.0$, and $\tau = 1.0$.
} 
\label{fig:sub} 
\end{figure}

\begin{figure}[ht]
\centering
{
    \includegraphics[width=6cm]{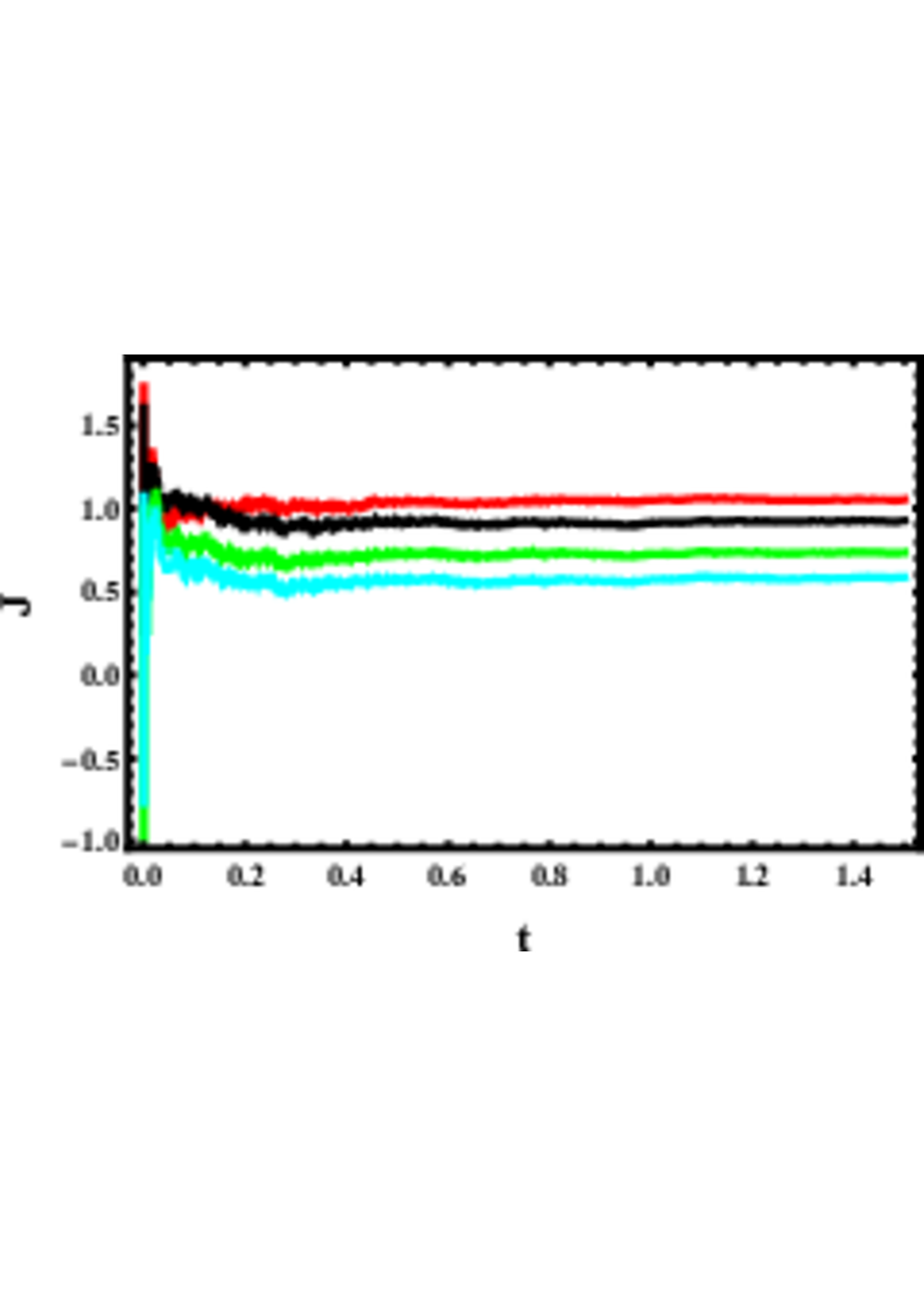}
}
\caption{ (Color online)  The plot of the current $J$ as a function of time is evaluated over a short time interval. The temperature is fixed at $\tau = 2$, whereas the potential $U_0$ is varied as $U_0=8$, $U_0=4$, $U_0=2$, and $U_0=1$ from top to bottom. Due to the non-uniform temperature, the current remains nonzero even when the load $\lambda = 0$.
 }
\label{fig:sub} 
\end{figure}

In Fig.  7a,  the trajectory $x(t)$ as a function of time $t$ is evaluated through numerical simulations for a fixed external force $f = 0.0$. The other parameters are held constant at $U_0 = 4.0$ and $\tau = 1.0$ (isothermal case). In Fig. 7b,  the probability distribution $P(x, t)$ as a function of time $t$ is analyzed under the same fixed parameters: $f = 0.0$, $U_0 = 4.0$, and $\tau = 1.0$. The probability of finding a particle is higher near the potential minima. However, as time progresses, this probability decreases, indicating that the entropy of the system increases over time.

\begin{figure}[ht]
\centering
{
    \includegraphics[width=6cm]{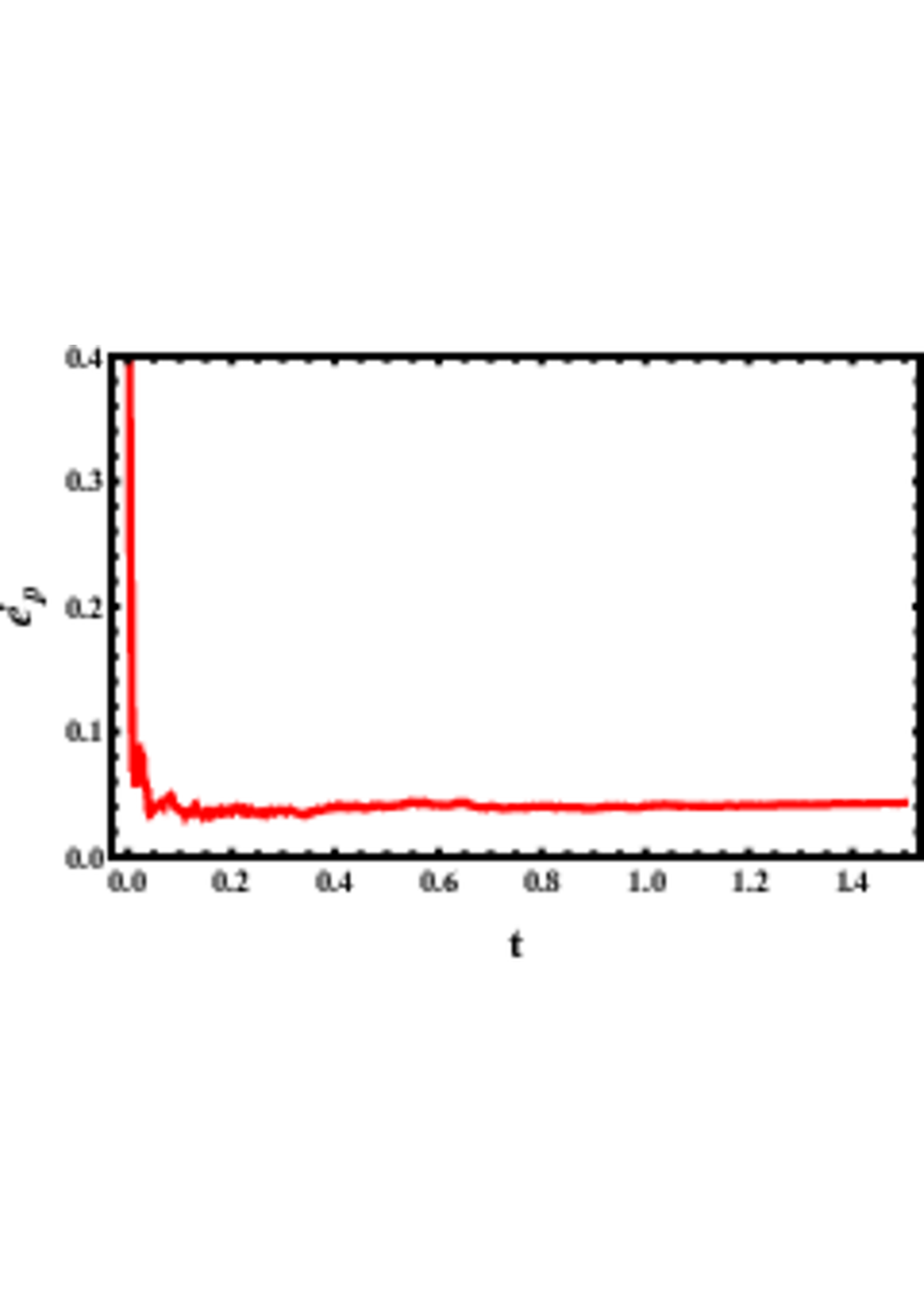}
}
\caption{ The entropy production rate  ($\dot{e}_p$) as a function of time $t$ for $U_0 = 1.0$, $\tau = 2.0$, and $\lambda = 0.0$ . 
} 
\end{figure}

The plot of the current $J$ as a function of time is evaluated over a short time interval, as shown in Fig. 8. In the figure, the temperature is fixed at $\tau = 2.0$, whereas the potential $U_0$ is varied as $U_0=8$, $U_0=4$, $U_0=2$, and $U_0=1$ from top to bottom. Due to the non-uniform temperature, the current remains nonzero even when the load $\lambda = 0$.
The rate of entropy production, $\dot{e}_p(t)$ is plotted in Fig. 9  for fixed parameter values of $U_0 = 1.0$, $\tau = 2.0$, and $\lambda = 0.0$. The figure shows that as time advances, the rate  decreases and approaches its steady-state value.

To summarize our findings, the analysis of short-time behavior through Brownian dynamics simulations demonstrates that under non-uniform temperature conditions, the particle undergoes a biased random walk, resulting in a nonzero velocity. The probability distribution of a particle's position evolves over time, signifying an increase in the entropy. The entropy production and extraction rates of the system adhere to fundamental thermodynamic principles, with entropy production initially dominating but decreasing over time. As the system transitions to a steady state, entropy production and extraction balance each other, ensuring a sustained nonequilibrium steady state.

\section{Summary and Conclusion}

We present a comprehensive analytical and numerical investigation of a Brownian heat engine operating at a periodic ratchet potential under an exponentially decreasing temperature profile. This thermal configuration closely models experimentally realizable gradients, such as those produced by laser-induced or plasmonic heating, and is directly relevant to energy conversion processes at the micro and nanoscale.

In the quasistatic limit, the engine achieves the exact Curzon–Ahlborn efficiency, $\eta = 1 - \sqrt{T_c/T_h}$, and its corresponding endoreversible coefficient of performance, $\mathrm{COP} = 1/\sqrt{T_h/T_c - 1}$, marking a rare case whichere these fundamental bounds emerge from a fully microscopic model. Brownian dynamics simulations confirm the analytical predictions and demonstrate how the current, velocity, and direction of transport depend on both the external load and thermal asymmetry. Depending on the operating conditions, the system functions either as a heat engine or as a refrigerator.

We further investigated networks of coupled Brownian motors and showed that while intensive quantities, such as the entropy production rate, remain invariant with system size, extensive quantities, such as the total entropy, scale linearly. Comparisons with linearly and piecewise decreasing thermal profiles reveal that the exponential gradient yields higher particle velocities and entropy production rates, albeit at the cost of lower thermodynamic efficiency, underscoring a fundamental trade-off between transport speed and energetic performance.

This work offers a solid theoretical framework that helps understand  how  Brownian heat engines operate  under realistic thermal conditions.  This study also  highlights the possible applications in nanoscale energy harvesting, directed heat transport, and thermally driven nanodevices. Thus, we introduce a model system that not only advances the understanding of thermodynamics in microscopic heat engines but also provides insight into the behavior of macroscopic endoreversible heat engines.

\section*{Acknowledgment}
I would like to thank  Mulu  Zebene  and Asfaw Taye for their
constant encouragement.

 \section*{Appendix I: Expression for the Particle Current}

To streamline notation and enhance readability, we introduce auxiliary symbols to represent frequently occurring terms:
\[
\kappa = \sqrt{T_h T_c}, \quad \theta = \sqrt{T_h/T_c}, \quad \Delta_{\pm} = fL \pm 2U_0, \quad \alpha = \ln(T_h / T_c).
\]

The steady-state particle current is given by
\begin{equation}
J = -\frac{\mathcal{F}}{\mathcal{G}_1 \mathcal{G}_2 + (\tau_1 + \tau_3)\mathcal{F}},
\end{equation}
where
\begin{widetext}
\begin{align}
\mathcal{F} &= -1 + \exp\left[
    \frac{(\kappa - T_h)\Delta_{-}}{T_c T_h \alpha} +
    \frac{(\theta - 1)\Delta_{+}}{T_h \alpha}
\right], \\[0.5em]
\mathcal{G}_1 &= g_1 + g_2, \\[0.5em]
g_1 &= \frac{L}{\Delta_{+}}
    \exp\left[-\frac{\Delta_{+}}{\kappa \alpha}\right]
    \left(
        \exp\left[\frac{\Delta_{+}}{\kappa \alpha}\right] -
        \exp\left[\frac{\Delta_{+}}{T_h \alpha}\right]
    \right), \\[0.5em]
g_2 &= \frac{L}{\Delta_{-}}
    \exp\left[
        \frac{fL T_c + 2 T_c U_0 - 4 U_0 \kappa}{T_c T_h \alpha}
    \right]
    \left(
        \exp\left[\frac{-\Delta_{-}}{\kappa \alpha}\right] -
        \exp\left[\frac{-2U_0 \ln T_c + 2U_0 \ln T_h - fL \alpha}{T_c \alpha^2}\right]
    \right), \\[0.5em]
\mathcal{G}_2 &= g_3 + g_4, \\[0.5em]
g_3 &= \frac{L}{\alpha}
    \exp\left[-\frac{\Delta_{+}}{T_h \alpha}\right]
    \left(
        \operatorname{Ei}\left[\frac{\Delta_{+}}{\kappa \alpha}\right] -
        \operatorname{Ei}\left[\frac{\Delta_{+}}{T_h \alpha}\right]
    \right), \\[0.5em]
g_4 &= \frac{L}{\alpha}
    \exp\left[\frac{-\Delta_{+} + 4 \theta U_0}{T_h \alpha}\right]
    \left(
        \operatorname{Ei}\left[\frac{\Delta_{-}}{T_c \alpha}\right] -
        \operatorname{Ei}\left[\frac{\Delta_{-}}{\kappa \alpha}\right]
    \right), \\[0.5em]
\tau_1 &= \frac{L^2}{2\Delta_{+}\alpha}
    \exp\left[-\frac{(1 + \theta)\Delta_{+}}{T_h \alpha}\right]
    \left[
        2 \exp\left[\frac{\Delta_{+}}{T_h \alpha}\right]
        \left(
            \operatorname{Ei}\left[\frac{\Delta_{+}}{T_h \alpha}\right] -
            \operatorname{Ei}\left[\frac{\Delta_{+}}{\kappa \alpha}\right]
        \right)
        - \exp\left[\frac{(1 + \theta)\Delta_{+}}{T_h \alpha}\right] \alpha
    \right], \\[0.5em]
\tau_3 &= \frac{L^2}{T_c \alpha^2}
    \exp\left[-\frac{4 U_0}{\kappa \alpha}\right]
    \left[
        \operatorname{Ei}\left[\frac{-\Delta_{-}}{T_c \alpha}\right] -
        \operatorname{Ei}\left[\frac{-\Delta_{-}}{\kappa \alpha}\right]
    \right]
    \left[
        \exp\left[\frac{4 U_0}{\kappa \alpha}\right]
        \left(
            \operatorname{Ei}\left[\frac{\Delta_{-}}{T_c \alpha}\right] -
            \operatorname{Ei}\left[\frac{\Delta_{-}}{\kappa \alpha}\right]
        \right)
        - \operatorname{Ei}\left[\frac{\Delta_{+}}{T_h \alpha}\right] +
        \operatorname{Ei}\left[\frac{\Delta_{+}}{\kappa \alpha}\right]
    \right].
\end{align}
\end{widetext}
\section*{Appendix II: Steady-State Probability Distribution for Exponentially Decreasing Temperature}

We derive the steady-state probability distribution \( P_s(x) \) for a Brownian particle moving in a periodic ratchet potential subjected to an external load \( f \). The total potential is given by
\[
U(x) = U_s(x) + f x,
\]
where \( U_s(x) \) is a piecewise linear ratchet potential and the temperature field is spatially varying as \( T(x) = T_h e^{-\alpha x} \).

The corresponding steady-state Fokker–Planck equation reads:
\begin{equation}
\frac{d}{dx} \left[ U'(x) P_s(x) + T(x) \frac{d P_s(x)}{dx} \right] = 0,
\end{equation}
where the drift and diffusion terms are given by \( A(x) = -\frac{dU(x)}{dx} \) and \( D(x) = \frac{k_B T(x)}{\gamma} \), respectively.

The potential profile leads to a piecewise constant drift term:
\begin{equation}
U'(x) =
\begin{cases}
 \frac{2U_0}{L_0} + f, & 0 < x \le \frac{L_0}{2}, \\
-\frac{2U_0}{L_0} + f, & \frac{L_0}{2} < x \le L_0.
\end{cases}
\end{equation}

In the steady-state regime with constant current, the Fokker–Planck equation simplifies to:
\begin{equation}
\frac{dP_s(x)}{dx} = -\frac{U'(x)}{D(x)} P_s(x),
\end{equation}
whose general solution can be written as
\begin{equation}
P_s(x) = B \frac{1}{T(x)} \exp\left(-\phi(x)\right), \quad \phi(x) = \int \frac{U'(x')}{k_B T(x')} dx'.
\end{equation}

Evaluating this expression for the exponential temperature profile \( T(x) = T_h e^{-\alpha x} \), we find in the region \( 0 < x \le L_0/2 \),
\begin{align}
\phi(x) &= \int \frac{-\left(\frac{2U_0}{L_0} + f\right)}{k_B T_h e^{-\alpha x}} dx = \frac{-\left(\frac{2U_0}{L_0} + f\right)}{k_B T_h \alpha} e^{\alpha x} + C_1,
\end{align}
and for \( L_0/2 < x \le L_0 \),
\begin{equation}
\phi(x) = \frac{\left(\frac{2U_0}{L_0} - f\right)}{k_B T_h \alpha} e^{\alpha x} + C_2.
\end{equation}

Continuity of the probability distribution at \( x = L_0/2 \) ensures a consistent relation between the integration constants \( C_1 \) and \( C_2 \). The full expression for the steady-state distribution is then obtained as follows.
\begin{widetext}
\begin{eqnarray}
P_s(x)& = &
\frac{B}{T_h} e^{\alpha x} \exp \left( 
    -\frac{\left(f + \frac{2U_0}{L_0}\right)L_0}{\alpha k_B T_h} 
    \left(e^{\alpha x/L_0} - 1 \right) 
\right), \quad 0 < x < \frac{L_0}{2}, \nonumber \\
&=& \frac{B}{T_h} e^{\alpha x} \exp \left( 
    -\frac{\left(f + \frac{2U_0}{L_0}\right)L_0}{\alpha k_B T_h} 
    \left(e^{\alpha / 2} - 1 \right) 
    - \frac{\left(f - \frac{2U_0}{L_0}\right)L_0}{\alpha k_B T_h} 
    \left(e^{\alpha x / L_0} - e^{\alpha / 2} \right) 
\right), \quad \frac{L_0}{2} < x < L_0.
\end{eqnarray}

In the limiting case where the ratchet potential vanishes (\( U_0 = 0 \)), the distribution simplifies to
\begin{equation}
P_s(x) = 
\frac{B}{T_h} e^{\alpha x} \exp \left( 
    -\frac{f L_0}{\alpha k_B T_h} 
    \left(e^{\alpha x/L_0} - 1 \right) 
\right).
\end{equation}
\end{widetext}
The normalization constant \( B \) is determined by the condition
\[
\int_0^{L_0} P_s(x)\, dx = 1.
\]


\begin{thebibliography}{60}

\bibitem{c8} P. H\"anggi, F. Marchesoni, and F. Nori, \textit{Ann. Phys. (Leipzig)} \textbf{14}, 51 (2005).

\bibitem{cc8} P. H\"anggi and F. Marchesoni, \textit{Rev. Mod. Phys.} \textbf{81}, 387 (2009).

\bibitem{mg1} A. Shi, H. Wu, and D. K. Schwartz, \textit{Sci. Adv.} \textbf{9}, eadj2208 (2023).

\bibitem{mg2} P. J. Zhang, G. K. Zhao, P. Wang, J. Huo, and X. M. Wang, \textit{Chaos Solitons Fract.} \textbf{171}, 113414 (2025).

\bibitem{mg3} B. Ai, \textit{Phys. Rev. E} \textbf{108}, 064409 (2023).

\bibitem{mg4} L. G. Chen, C. Z. Qi, Y. L. Ge, and H. J. Feng, \textit{Energy} \textbf{255}, 124582 (2022).

\bibitem{mg5} L. Lin, L. Yu, W. Lv, and H. Wang, \textit{Chin. J. Phys.} \textbf{68}, 78 (2020).

\bibitem{mg6} M. Wi\'sniewski and J. Spiechowicz, \textit{Phys. Rev. E} \textbf{111}, 024130 (2025).

\bibitem{mg7} M. Patel and D. Chaudhuri, arXiv:2310.00802 [cond-mat.stat-mech].

\bibitem{mg8} M. Sanoria, R. Chelakkot, and A. Nandi, \textit{Soft Matter} \textbf{20}, 9184 (2024).

\bibitem{am7} P. Reimann, R. Bartussek, R. H\"aussler, and P. H\"anggi, \textit{Phys. Lett. A} \textbf{215}, 26 (1996).

\bibitem{c9} M. A. Taye and M. Bekele, \textit{Eur. Phys. J. B} \textbf{38}, 457 (2004).

\bibitem{c10} M. A. Taye and M. Bekele, \textit{Phys. Rev. E} \textbf{72}, 056109 (2005).

\bibitem{c11} M. A. Taye and M. Bekele, \textit{Physica A} \textbf{384}, 346 (2007).

\bibitem{c12} M. Matsuo and S. Sasa, \textit{Physica A} \textbf{276}, 188 (1999).

\bibitem{c13} I. Der\'enyi and R. D. Astumian, \textit{Phys. Rev. E} \textbf{59}, R6219 (1999).

\bibitem{c14} I. Der\'enyi, M. Bier, and R. D. Astumian, \textit{Phys. Rev. Lett.} \textbf{83}, 903 (1999).

\bibitem{c15} J. M. Sancho, M. S. Miguel, and D. D\"urr, \textit{J. Stat. Phys.} \textbf{28}, 291 (1982).

\bibitem{c16} B. Q. Ai, H. Z. Xie, D. H. Wen, X. M. Liu, and L. G. Liu, \textit{Eur. Phys. J. B} \textbf{48}, 101 (2005).

\bibitem{c17} M. A. Taye, \textit{Eur. Phys. J. B} \textbf{86}, 189 (2013).

\bibitem{c18} F. L. Curzon and B. Ahlborn, \textit{Am. J. Phys.} \textbf{43}, 22 (1975).

\bibitem{c19} M. A. Taye, \textit{Phys. Rev. E} \textbf{89}, 012143 (2014).

\bibitem{c20} M. A. Taye, \textit{Phys. Rev. E} \textbf{94}, 032111 (2016).

\bibitem{c21} M. A. Taye, \textit{Phys. Rev. E} \textbf{101}, 012131 (2020).

\bibitem{c23} M. A. Taye, \textit{Phys. Rev. E} \textbf{105}, 054126 (2022).

\bibitem{c22} M. A. Taye, \textit{Phys. Rev. E} \textbf{110}, 054105 (2024).

\bibitem{c25} S. Rekhi, J. Tempere, and I. F. Silvera, \textit{Rev. Sci. Instrum.} \textbf{74}, 3820 (2003).

\bibitem{c26} D. Braun and A. Libchaber, \textit{Phys. Rev. Lett.} \textbf{89}, 188103 (2002).

\bibitem{c27} E. J. G. Peterman, F. Gittes, and C. F. Schmidt, \textit{Biophys. J.} \textbf{84}, 1308 (2003).

\bibitem{c28} P. Yu and Y. Zeng, \textit{Int. J. Heat Mass Transfer} \textbf{106}, 989 (2017).

\end{thebibliography}
\end{document}